\documentclass[12pt,fleqn]{article}
\usepackage{mcite,amsfonts}
\usepackage{style}
\usepackage{diagrams}
\setdefaultlengths
\begin{document}
\begin{titlepage}
\pub{10}{96}
\title{${\cal O}(\alpha_s^2)$ Contributions to the asymmetric fragmentation
  function in $e^+e^-$ annihilation}
    {P.J. Rijken and W.L. van Neerven}
    {August 1996}
\abstract{
The order \alphastwo~contributions to the coefficient functions corresponding to
the asymmetric fragmentation function $F_A(x,Q^2)$ in $e^+e^-$ annihilation are calculated.
From this calculation we infer that the order $(\alpha_s/4\pi)^2$ correction to the flavour
asymmetry sum rule is non vanishing and amounts to $-12\beta_0C_F\zeta(3)$. We also
study the effect of the higher order QCD corrections on $F_A(x,Q^2)$ and compare them with
the OPAL data. The latter put a strong constraint on the valence part of the
fragmentation densities $D_q^H(x,\mu^2)$.
}
\end{titlepage}
%

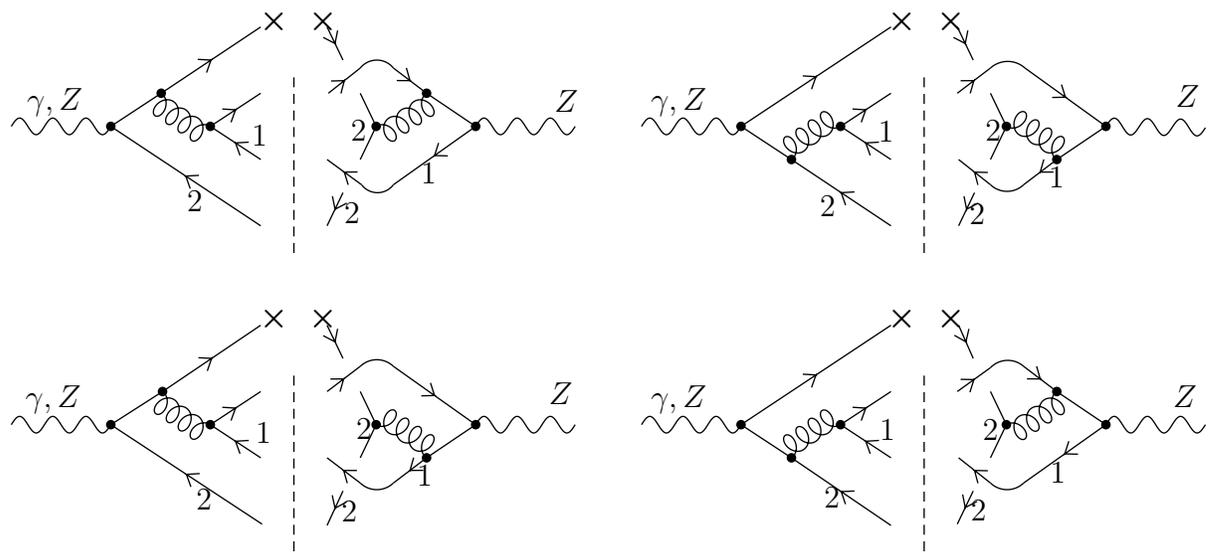
\begin{figure}
  \begin{center}
    \setlength{\unitlength}{25pt}
\renewcommand{\Init}{0.500 0.500 Scale Init}
\begin{picture}(19.000,10.000)
\put(0,0){\VectorBoson{1.000}{15.000}{4.000}{15.000}{}}
\put(0,0){\VectorBoson{15.000}{15.000}{18.000}{15.000}{}}
\put(0,0){\Fermion{8.520}{12.000}{4.000}{15.000}{+}}
\put(0,0){\Fermion{5.520}{16.000}{8.520}{18.000}{+}}
\put(0,0){\Fermion{7.000}{15.000}{8.520}{16.000}{+}}
\put(0,0){\Fermion{8.520}{14.000}{7.000}{15.000}{+}}
\put(0,0){\Fermion{12.520}{16.760}{13.520}{16.000}{+}}
\put(0,0){\Fermion{15.000}{15.000}{12.520}{13.240}{+}}
\put(0,0){\Fermion{11.560}{13.240}{10.520}{14.000}{+}}
\put(0,0){\Fermion{10.560}{16.000}{11.520}{16.760}{+}}
\put(0,0){\Fermion{11.000}{13.000}{10.520}{12.000}{+}}
\put(0,0){\Fermion{10.520}{18.000}{11.000}{17.000}{+}}
\put(0,0){\Fermion{4.000}{15.000}{5.520}{16.000}{}}
\put(0,0){\Fermion{11.520}{16.000}{12.000}{15.000}{}}
\put(0,0){\Fermion{12.000}{15.000}{11.520}{14.000}{}}
\put(0,0){\Fermion{13.520}{16.000}{15.000}{15.000}{}}
\put(0,0){\Gluon{5.520}{16.000}{7.000}{15.000}{}}
\put(0,0){\Gluon{12.000}{15.000}{13.520}{16.000}{}}
\put(0,0){\FermionArc{11.560}{13.240}{12.520}{13.240}{-106.260}{}}
\put(0,0){\FermionArc{11.520}{16.760}{12.520}{16.760}{102.564}{}}
\put(0,0){\ScaGho{9.520}{16.480}{9.520}{11.000}{}}
\put(0,0){\VectorBoson{20.000}{15.000}{23.000}{15.000}{}}
\put(0,0){\VectorBoson{34.000}{15.000}{37.000}{14.960}{}}
\put(0,0){\Fermion{23.000}{15.000}{27.520}{18.000}{+}}
\put(0,0){\Fermion{29.520}{18.000}{30.000}{17.000}{+}}
\put(0,0){\Fermion{31.560}{16.760}{34.000}{15.000}{+}}
\put(0,0){\Fermion{32.520}{14.000}{31.520}{13.240}{+}}
\put(0,0){\Fermion{30.520}{13.240}{29.560}{14.000}{+}}
\put(0,0){\Fermion{29.520}{16.000}{30.520}{16.760}{+}}
\put(0,0){\Fermion{26.000}{15.000}{27.520}{16.000}{+}}
\put(0,0){\Fermion{27.520}{14.000}{26.000}{15.000}{+}}
\put(0,0){\Fermion{27.520}{12.000}{24.520}{14.000}{+}}
\put(0,0){\Fermion{30.040}{13.000}{29.560}{12.000}{+}}
\put(0,0){\Fermion{23.000}{15.000}{24.520}{14.000}{}}
\put(0,0){\Fermion{34.000}{15.000}{32.520}{14.000}{}}
\put(0,0){\Fermion{30.520}{16.000}{31.000}{15.000}{}}
\put(0,0){\Fermion{31.000}{15.000}{30.520}{14.000}{}}
\put(0,0){\FermionArc{30.520}{16.760}{31.560}{16.760}{84.150}{}}
\put(0,0){\FermionArc{31.520}{13.240}{30.520}{13.240}{87.206}{}}
\put(0,0){\ScaGho{28.520}{16.480}{28.520}{11.000}{}}
\put(0,0){\Gluon{26.000}{15.000}{24.520}{14.000}{}}
\put(0,0){\Gluon{32.520}{14.000}{31.000}{15.000}{}}
\put(0,0){\VectorBoson{1.000}{6.000}{4.000}{6.000}{}}
\put(0,0){\VectorBoson{15.000}{6.000}{18.000}{6.000}{}}
\put(0,0){\VectorBoson{20.000}{6.000}{23.000}{6.000}{}}
\put(0,0){\VectorBoson{34.000}{6.000}{37.000}{6.000}{}}
\put(0,0){\Fermion{5.560}{7.000}{8.520}{9.000}{+}}
\put(0,0){\Fermion{8.560}{3.000}{4.000}{6.000}{+}}
\put(0,0){\Fermion{7.000}{6.000}{8.520}{7.000}{+}}
\put(0,0){\Fermion{8.520}{5.000}{7.000}{6.000}{+}}
\put(0,0){\Fermion{12.520}{7.760}{15.000}{6.000}{+}}
\put(0,0){\Fermion{13.520}{5.000}{12.520}{4.240}{+}}
\put(0,0){\Fermion{11.520}{4.240}{10.520}{5.000}{+}}
\put(0,0){\Fermion{11.000}{4.000}{10.520}{2.960}{+}}
\put(0,0){\Fermion{10.520}{9.000}{11.000}{8.000}{+}}
\put(0,0){\Fermion{10.520}{7.000}{11.520}{7.760}{+}}
\put(0,0){\Fermion{4.000}{6.000}{5.560}{7.000}{}}
\put(0,0){\Fermion{15.000}{6.000}{13.520}{5.000}{}}
\put(0,0){\Fermion{12.000}{6.000}{11.520}{5.000}{}}
\put(0,0){\Fermion{11.520}{7.000}{12.000}{6.000}{}}
\put(0,0){\FermionArc{11.520}{7.760}{12.520}{7.760}{87.206}{}}
\put(0,0){\FermionArc{12.520}{4.240}{11.520}{4.240}{87.206}{}}
\put(0,0){\Gluon{5.560}{7.000}{7.000}{6.000}{}}
\put(0,0){\Gluon{13.520}{5.000}{12.000}{6.000}{}}
\put(0,0){\ScaGho{9.520}{7.480}{9.520}{2.000}{}}
\put(0,0){\Fermion{23.000}{6.000}{27.520}{9.000}{+}}
\put(0,0){\Fermion{27.520}{2.960}{24.520}{5.000}{+}}
\put(0,0){\Fermion{26.000}{6.000}{27.520}{7.000}{+}}
\put(0,0){\Fermion{27.520}{5.000}{26.000}{6.000}{+}}
\put(0,0){\Fermion{29.560}{9.000}{30.000}{8.000}{+}}
\put(0,0){\Fermion{30.000}{4.000}{29.520}{3.000}{+}}
\put(0,0){\Fermion{34.000}{6.000}{31.520}{4.240}{+}}
\put(0,0){\Fermion{30.520}{4.240}{29.560}{5.000}{+}}
\put(0,0){\Fermion{29.520}{7.000}{30.520}{7.720}{+}}
\put(0,0){\Fermion{23.000}{6.000}{24.520}{5.000}{}}
\put(0,0){\Fermion{30.520}{7.000}{31.000}{6.000}{}}
\put(0,0){\Fermion{31.000}{6.000}{30.520}{5.000}{}}
\put(0,0){\Fermion{31.520}{7.760}{32.520}{7.000}{+}}
\put(0,0){\Fermion{32.520}{7.000}{34.000}{6.000}{}}
\put(0,0){\Gluon{26.000}{6.000}{24.520}{5.000}{}}
\put(0,0){\Gluon{31.000}{6.000}{32.520}{7.000}{}}
\put(0,0){\FermionArc{30.520}{7.720}{31.520}{7.760}{87.079}{}}
\put(0,0){\FermionArc{31.520}{4.240}{30.520}{4.240}{87.206}{}}
\put(0,0){\ScaGho{28.520}{7.480}{28.520}{2.000}{}}
\put(0,0){\Bullet{4.000}{15.000}}
\put(0,0){\Bullet{5.520}{16.000}}
\put(0,0){\Bullet{7.000}{15.000}}
\put(0,0){\Bullet{12.000}{15.000}}
\put(0,0){\Bullet{15.000}{15.000}}
\put(0,0){\Bullet{13.520}{16.000}}
\put(0,0){\Bullet{15.000}{6.000}}
\put(0,0){\Bullet{13.520}{5.000}}
\put(0,0){\Bullet{12.000}{6.000}}
\put(0,0){\Bullet{7.000}{6.000}}
\put(0,0){\Bullet{5.560}{7.000}}
\put(0,0){\Bullet{4.000}{6.000}}
\put(0,0){\Bullet{26.000}{15.000}}
\put(0,0){\Bullet{24.520}{14.000}}
\put(0,0){\Bullet{23.000}{15.000}}
\put(0,0){\Bullet{31.000}{15.000}}
\put(0,0){\Bullet{32.520}{14.000}}
\put(0,0){\Bullet{34.000}{15.000}}
\put(0,0){\Bullet{34.000}{6.000}}
\put(0,0){\Bullet{32.520}{7.000}}
\put(0,0){\Bullet{31.000}{6.000}}
\put(0,0){\Bullet{26.000}{6.000}}
\put(0,0){\Bullet{24.520}{5.000}}
\put(0,0){\Bullet{23.000}{6.000}}
\put(4.260,9.000){\boldmath$\times$}
\put(5.000,9.000){\boldmath$\times$}
\put(4.260,4.500){\boldmath$\times$}
\put(5.000,4.500){\boldmath$\times$}
\put(13.760,9.000){\boldmath$\times$}
\put(14.500,9.000){\boldmath$\times$}
\put(13.760,4.500){\boldmath$\times$}
\put(14.500,4.500){\boldmath$\times$}
\put(4.140,7.240){1}
\put(6.700,6.660){1}
\put(13.620,7.300){1}
\put(16.200,6.600){1}
\put(4.200,2.740){1}
\put(6.640,2.080){1}
\put(13.640,2.780){1}
\put(16.220,2.140){1}
\put(3.160,6.240){2}
\put(5.540,6.060){2}
\put(5.640,7.320){2}
\put(12.740,6.160){2}
\put(15.000,6.040){2}
\put(15.240,7.320){2}
\put(3.300,1.720){2}
\put(5.500,1.540){2}
\put(5.720,2.800){2}
\put(12.800,1.760){2}
\put(15.020,1.560){2}
\put(15.200,2.780){2}
\put(0.740,7.780){$\gamma,Z$}
\put(0.720,3.280){$\gamma,Z$}
\put(8.720,7.740){$Z$}
\put(8.640,3.320){$Z$}
\put(10.180,7.780){$\gamma,Z$}
\put(10.180,3.280){$\gamma,Z$}
\put(18.120,7.820){$Z$}
\put(18.080,3.280){$Z$}
\end{picture}
    \caption{\label{fig:1}}
  \end{center}
\end{figure}

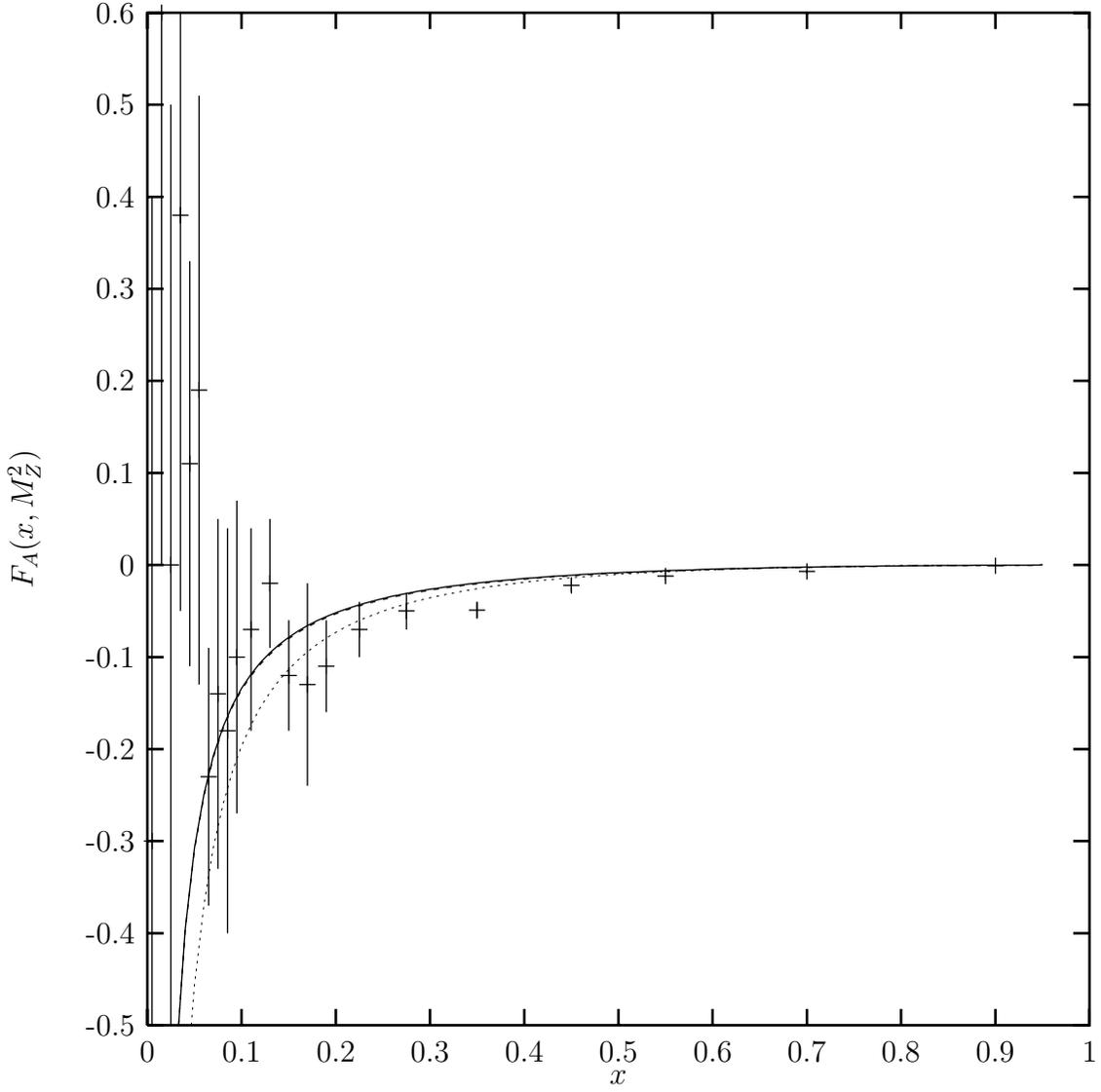
\begin{figure}
  \begin{center}
    \input figure1
    \caption{\label{fig:2}}
  \end{center}
\end{figure}

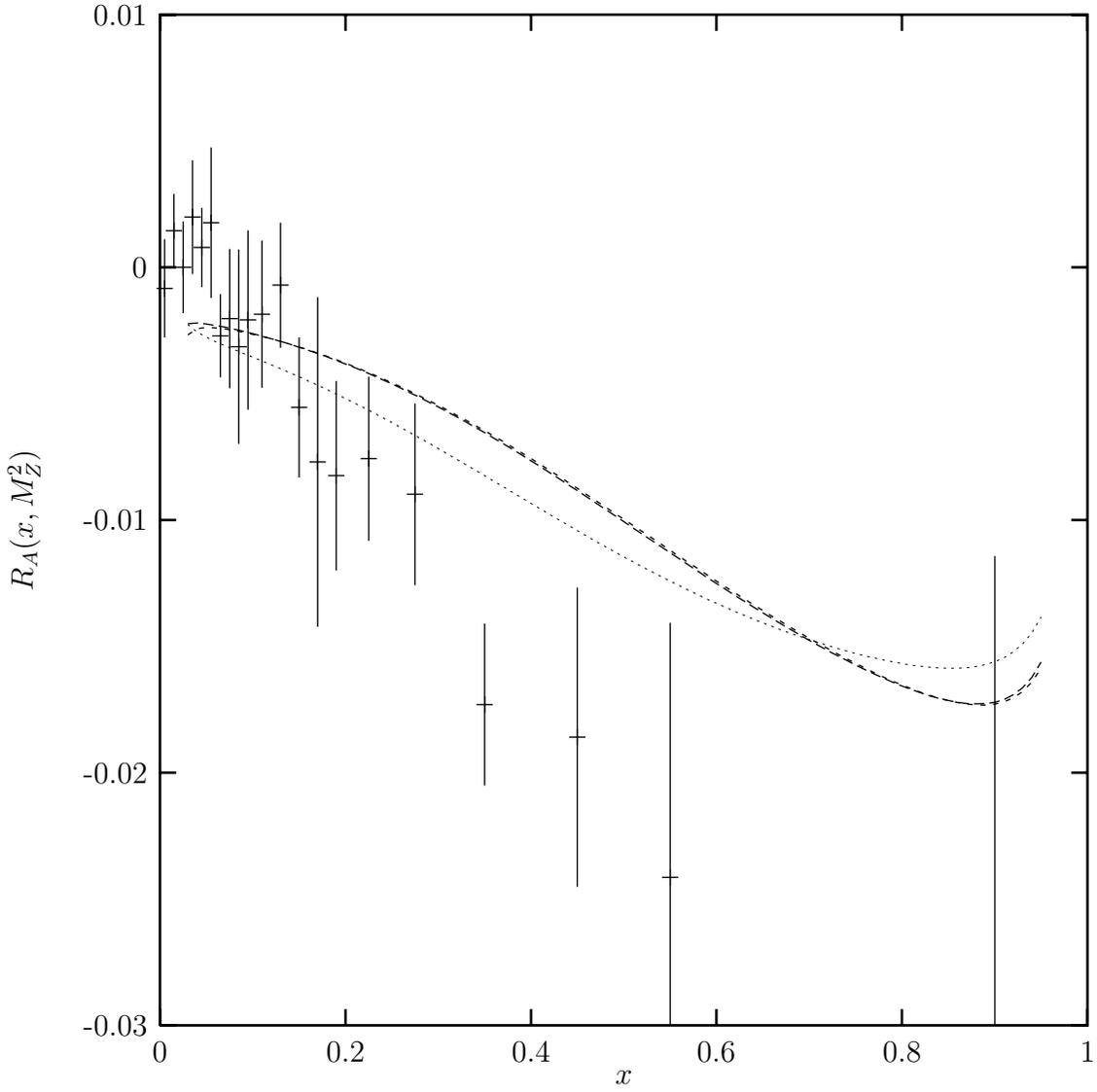
\begin{figure}
  \begin{center}
    \input figure2
    \caption{\label{fig:3}}
  \end{center}
\end{figure}

\newpage
\noindent
The measurement of the fragmentation functions in the process
\begin{equation}
  \label{eq:1}
  e^+ e^- \rightarrow \gamma,Z \rightarrow H + ``X",
\end{equation}
provides us in addition to other experiments like deep inelastic lepton-hadron
scattering, with a new test of scaling violation as predicted by perturbative
quantum chromodynamics (QCD). Here $``X"$ denotes any inclusive hadronic
final state and $H$ represents either a specific charged outgoing hadron or
a sum over all charged hadron species. This process has been studied over a
wide range of energies of many different $e^+e^-$-colliders. Data haven been
collected from TASSO \cite{Bra90} ($\sqrt{s}=22, 35, 45$ GeV), MARK II \cite{Pet88}
and TPC/2$\gamma$ \cite{Aih88} ($\sqrt{s}=29$ GeV), CELLO \cite{Pod}
($\sqrt{s}=35$ GeV), AMY \cite{Li90} ($\sqrt{s}=55$ GeV) and DELPHI
\cite{Abr93}, ALEPH \cite{Bus95}, OPAL \cite{Ake95} ($\sqrt{s}=91.2$ GeV).\\
Following the notation in \cite{Nas94} the unpolarized differential cross
section of process \eref{eq:1} is given by
\begin{equation}
  \label{eq:2}
  \frac{d^2\sigma^H}{dx\,d\cos\theta} = \frac{3}{8}(1+\cos^2\theta)\frac{d\sigma_T^H}{dx}
  + \frac{3}{4}\sin^2\theta\frac{d\sigma_L^H}{dx} + \frac{3}{4}\cos\theta
  \frac{d\sigma_A^H}{dx},
\end{equation}
where $x$ stands for the Bj{\o}rken scaling variable
\begin{equation}
  \label{eq:3}
  x = \frac{2pq}{Q^2},\hspace*{8mm} 0<x\leq 1,
\end{equation}
and $p$ and $q$ ($q^2=Q^2>0$) are the four-momenta of the produced particle $H$
and the virtual vector boson ($\gamma,Z$) respectively. The variable $\theta$
denotes the angle of emission of particle $H$ with respect to the electron beam
direction in the center of mass (CM) frame. The transverse, longitudinal and
asymmetric cross section in \eref{eq:2} are defined by $\sigma_T,\sigma_L$ and
$\sigma_A$ respectively. The latter only shows up if the intermediate vector
boson is given by the $Z$-boson and is absent in purely electromagnetic
annihilation. Before the advent  of LEP1 the CM energies were so low
($\sqrt{s} < M_Z$) that $\sigma_A$ could not be measured and no effort was made
to separate $\sigma_L$ from $\sigma_T$ so that only data for
$d^2\sigma^H/dxd\cos\theta$ \eref{eq:2} were available. Recently, after LEP1 came
into operation. ALEPH \cite{Bus95} and OPAL \cite{Ake95} obtained data for
$\sigma_L$ and $\sigma_T$ separately and the latter collaboration even made a
measurement of $\sigma_A$ for the first time. The separation of $\sigma_L$
and $\sigma_T$ is important because the former cross section enables us to
extract the strong coupling constant \alphas~and allows us to determine
the gluon fragmentation density $D_g(x)$ with a much higher degree of accuracy
as could be done before. Furthermore the measurement of $\sigma_A$ provides us
with information on hadronization effects \cite{Nas94} since the QCD corrections
are very small.\\
As far as the theoretical achievements are concerned the order \alphas~QCD
corrections to the coefficient functions $\mathbb{C}_{k,\ell}$ ($k=T,L,A$,
$\ell=q,g$), corresponding to the cross sections $\sigma_k$ \eref{eq:2}, have
been calculated in the past in \cite{Bai79,Alt79b} (see also \cite{Nas94}). Also
computed are the NLO corrections to the DGLAP timelike splitting functions
$P_{ij}(x)$ ($i,j=q,g$) in \cite{Cur80,*Fur80,*Flo81}. Recently the order
\alphastwo~contributions to the coefficient functions $\mathbb{C}_{L,i}$
\cite{Rij96} and $\mathbb{C}_{T,i}$ \cite{Rij96b} became available. From the
latter one obtains the order \alphastwo~corrections to the total longitudinal
and transverse cross sections defined by
\begin{equation}
  \label{eq:4}
  \sigma_k(Q^2) = \frac{1}{2}\sum_H\,\int_0^1dx\,\,x\frac{d\sigma_k^H(x,Q^2)}{dx},
\end{equation}
which are equal to ($k=T,L$)
\begin{equation}
  \label{eq:5}
  \sigma_k(Q^2) = \sigma^{(0)}(Q^2)\,\int_0^1dz\,\,z\BigLeftHook
  \mathbb{C}_{k,q}^S(z,Q^2/M^2) + \frac{1}{2}\mathbb{C}_{k,g}(z,Q^2/M^2)
  \BigRightHook.
\end{equation}
Here $\sigma^{(0)}(Q^2)$ is the zeroth order annihilation cross section of
process \eref{eq:1} which is identical to $\sum_f\sigma_{0,f}$ in eq. (2.14)
of \cite{Nas94} where $f$ denotes a specific flavour ($f=u,d,s,c,b$). Furthermore
$\sigma_{\rm tot}(e^+e^-\rightarrow X) = \sigma_T + \sigma_L$ which has been
calculated up to order $\alpha_s^3$ in \cite{Gor91,*Sur91}. In \cite{Rij96,Rij96b}
it was shown that the order \alphastwo~contributions to the coefficient functions
are necessary to get agreement between the
OPAL-data and the theoretical predictions. Since the OPAL-collaboration also measured
the asymmetric cross section $d\sigma_A^H/dx$ in \eref{eq:2} it will be of interest to
compute the order \alphastwo~corrections to this quantity too. In the QCD
improved parton model it can be written as
\begin{eqnarray}
  \label{eq:6}
  \lefteqn{
    \frac{d\sigma_A^H}{dx}(x,Q^2) = \sum_f\,\int_x^1\frac{dz}{z}\,A_f(Q^2)
    \BigLeftParen D_f^H\left(\frac{x}{z},\mu^2\right)
    - D_{\bar{f}}^H\left(\frac{x}{z},\mu^2\right)\BigRightParen\,\mathbb{C}_{A,q}
    ^\NonSinglet(z,Q^2/\mu^2),}\nonumber\\[2ex]
  &&
\end{eqnarray}
where $f$ denotes the flavour of the quark.
The asymmetry factor $A_f$, which is defined in eq. (2.12) of \cite{Nas94},
contains the products of the vector and axial vector electroweak couplings
appearing in the $\gamma-Z$ and $Z-Z$ interference term. It also includes
the contribution of the photon and $Z$-boson propagators. The parton
fragmentation densities denoted by $D_f^H(z,\mu^2)$ depend in addition to
the partonic scaling variable $z$ also on the factorization scale $\mu$. Because
of charge conjugation symmetry of the strong interactions we have used in
\eref{eq:6} the identities
\begin{equation}
  \label{eq:7}
  \mathbb{C}_{A,f}^\NonSinglet = -\mathbb{C}_{A,\bar{f}}^\NonSinglet \equiv
  \mathbb{C}_{A,q}^\NonSinglet,\hspace*{8mm}
  \mathbb{C}_{A,q}^\Singlet = \mathbb{C}_{A,g}^\NonSinglet = 0.
\end{equation}
Up to order \alphastwo~ the non-singlet coefficient function $\mathbb{C}_{A,q}^\NonSinglet$
receives contributions from the following parton subprocesses where all quarks
are taken to be massless
\begin{eqnarray}
  \label{eq:8}
  {\cal O}(\alpha_s^0) &:& V \rightarrow ``q" + \bar{q},\\[2ex]
  \label{eq:9}
  {\cal O}(\alpha_s)   &:& V \rightarrow ``q" + \bar{q} + g,\\[2ex]
  \label{eq:10}
  {\cal O}(\alpha_s^2) &:& V \rightarrow ``q" + \bar{q} + g + g,\\[2ex]
  \label{eq:11}
                       &&  V \rightarrow ``q" + \bar{q} + q + \bar{q}.
\end{eqnarray}
Here the detected quark, which fragments into the hadron, is indicated by
$``q"$ and $V=\gamma,Z$. The above reactions also include the one- and
two-loop corrections to process \eref{eq:8} and the one-loop corrections to
process \eref{eq:9}. Further in reaction \eref{eq:11} the two anti-quarks can
be identical as well as non-identical. In the case the anti-quark is detected
one has to interchange $q$ and $\bar{q}$ in eqs. \eref{eq:8}-\eref{eq:11}.\\
The computation of the parton cross sections proceeds in the same way as has been
done for the longitudinal $\mathbb{C}_{L,i}$ and transverse coefficient
functions $\mathbb{C}_{T,i}$ presented in \cite{Rij96} and \cite{Rij96b}
respectively. In the calculation one has to deal with the presence of ultraviolet
(UV), infrared (IR) and collinear (C) divergences which have to be regularized
using the method of $n$-dimensional regularization. However there is one
difference between the calculation of $\mathbb{C}_{k,i}$ ($k=T,L$) on one hand
and the computation of $\mathbb{C}_{A,i}$ on the other hand. This difference
is due to the appearance of the $\gamma_5$-matrix in the interference term
$M_VM_A^\ast + M_AM_V^\ast$ where $M_V$ and $M_A$ stand for the vector and
axial-vector amplitude of the above processes. Here one has to find an
$n$-dimensional extension for the $\gamma_5$-matrix occuring in $M_A$. For our
calculation we have adopted the prescription for $\gamma_5$ given by 't Hooft
and Veltman \cite{Hoo72} (see also Breitenlohner and Maison \cite{Bre77}). Since
the axial vector vertex is represented by $\gamma_\mu\gamma_5$ one can simplify
the traces using the identification
\begin{equation}
  \label{eq:12}
  \gamma_\mu\gamma_5 = -\frac{i}{6}\epsilon_{\mu\alpha\beta\sigma}
  \gamma^\alpha\gamma^\beta\gamma^\sigma,
\end{equation}
which yields the same result as the prescription of 't Hooft and Veltman as is
shown in \cite{Aky73,-Aky73b,-Aky74,Lar93,-Lar94c}. Although this prescription is consistent it has
one drawback namely that the non-singlet axial vector current is renormalized
in spite of the fact that it is conserved. Hence for each virtual correction
where the $\gamma_5$-matrix appears in the loop one needs an additional
renormalization constant to undo this unwanted effect. This constant has been
calculated in \cite{Lar93,-Lar94c} and reads up to order \alphastwo
\begin{eqnarray}
  \label{eq:13}
  \lefteqn{
    Z_A = 1 - \frac{\alpha_s}{4\pi}C_F\BigLeftHook 4 - 5\varepsilon\BigRightHook
    + \left(\frac{\alpha_s}{4\pi}\right)\BigLeftHook C_F^2\BigLeftBrace 22
    \BigRightBrace + C_AC_F\BigLeftBrace -\frac{44}{3\varepsilon} - \frac{107}{9}}
  \nonumber\\[2ex]
  && {}+ n_fC_FT_f\BigLeftBrace\frac{16}{3\varepsilon}+\frac{4}{9}\BigRightBrace,
\end{eqnarray}
where the colour factors in QCD are given by $C_F = (N^2-1)/2N$, $C_A=N$, and $T_f=1/2$
with $N=3$
and the number of light flavours is denoted by $n_f$. The rest of the calculation
proceeds in the same way as performed for the deep inelastic coefficient
functions $\mathbb{C}_{3,q}^\NonSinglet$ \cite{Zij92} which is the analogue of
$\mathbb{C}_{A,q}^\NonSinglet$. Apart from the check on the procedure outlined in
\cite{Zij92} we will add a new one which has the advantage that we can get rid
of the renormalization constant $Z_A$ in \eref{eq:13} which is needed when the
$\gamma_5$-matrix appears in the loop of the virtual Feynman graph. This check
is based on the observation that the difference between the asymmetric and
transverse parton cross sections does not contain the distributions denoted by
$\delta(1-z)$ and $(\ln^k(1-z)/(1-z))_+$. These singular functions originate
from the one- and two-loop corrections to the Born-process \eref{eq:8} and
the contributions due to soft gluon and collinear fermion pair production in reactions
\eref{eq:9}-\eref{eq:11}. Hence these distributions cancel in
$\mathbb{C}_{A,q}^\NonSinglet(z,Q^2/\mu^2)-\mathbb{C}_{T,q}^\NonSinglet(z,Q^2/\mu^2)$. Since
the transverse coefficient function $\mathbb{C}_{T,q}^\NonSinglet$ is known \cite{Rij96b}
we can obtain $\mathbb{C}_{A,q}^\NonSinglet$ from the difference
$\mathbb{C}_{A,q}^\NonSinglet-\mathbb{C}_{T,q}^\NonSinglet$. The latter is only determined
by the one-loop corrections to the regular part of process \eref{eq:9}
(hard gluon radiative part) and the regular part of \eref{eq:10}, \eref{eq:11}
(hard gluon radiation plus quark anti-quark production). Hence we only have to deal
with the $\gamma_5$-matrix
in the one-loop corrections to \eref{eq:9}. However we have now checked that the
following identity holds
\begin{eqnarray}
  \label{eq:14}
  \lefteqn{
    Z_A\BigLeftHook\BigLeftBrace {M_V^{(1)}}^\ast M_A^{(1)} + {M_A^{(1)}}^\ast M_V^{(1)}
    \BigRightBrace + \BigLeftBrace {M_V^{(1)}}^\ast M_A^{(3)} + {M_V^{(3)}}^\ast M_A^{(1)}
    + {M_A^{(1)}}^\ast M_V^{(3)}} \nonumber\\[2ex]
  && {}+ {M_A^{(3)}}^\ast M_V^{(1)} \BigRightBrace\BigRightHook =
  \BigLeftBrace {M_V^{(1)}}^\ast M_A^{(1)} + {M_A^{(1)}}^\ast M_V^{(1)}\BigRightBrace
  + \BigLeftBrace 2{M_A^{(1)}}^\ast M_V^{(3)} \nonumber\\[2ex]
  && {}+ 2{M_V^{(3)}}^\ast M_A^{(1)}\BigRightBrace,
\end{eqnarray}
where $M_k^{(\ell)}$ is the order $g^\ell$ ($\alpha_s = g^2/4\pi$) contribution to the
amplitude $M_k$ ($k=V,A$). Here $M_V^{(\ell)}$ and $M_A^{(\ell)}$ denote the amplitudes
where the quark is attached to the vector- and axial-vector current respectively so that
$M_A^{(\ell)}$ contains the $\gamma_5$-matrix. The term between the first pair of curly
brackets on the left-hand
side of \eref{eq:14} originates from the purely radiative process \eref{eq:9} whereas the term
in the second pair of curly brackets refers to the interference between process \eref{eq:9}
and the virtual corrections to \eref{eq:9}. The latter is represented by the amplitude
$M_k^{(3)}$ ($k=V,A$). Equation \eref{eq:14} reveals that one can get rid of the
renormalization constant $Z_A$ by shifting the $\gamma_5$-matrix from $M_A^{(3)}$
to the amplitude $M_A^{(1)}$ of the radiative process \eref{eq:9}
so that this matrix becomes harmless.
Actually one can now also choose the naive $\gamma_5$-prescription without altering
the final result. The order \alphas~corrections to $\mathbb{C}_{T,q}^\NonSinglet$ are
calculated in \cite{Alt79b} (see also eqs. (2.15), (2.16) in \cite{Nas94}). The
order \alphastwo~corrections calculated in this paper are presented as follows.
First we split the coefficient functions $\mathbb{C}_{k,q}^\NonSinglet$ ($k=T,~A$)
in two parts i.e.
\begin{eqnarray}
  \label{eq:pre15}
  \lefteqn{\mathbb{C}_{T,q}^\NonSinglet = \mathbb{C}_{T,q}^{\NonSinglet,{\rm nid}}
  + \mathbb{C}_{T,q}^{\NonSinglet,{\rm id}},}\\[2ex]
  \label{eq:15}
  \lefteqn{\mathbb{C}_{A,q}^\NonSinglet = \mathbb{C}_{A,q}^{\NonSinglet,{\rm nid}}
  - \mathbb{C}_{A,q}^{\NonSinglet,{\rm id}},}
\end{eqnarray}
where the last term in the above equations is only due to the contribution from identical
anti-quarks in reaction \eref{eq:11}. The second order contributions to both parts can
be now obtained from $\mathbb{C}_{T,q}^{\NonSinglet,{\rm nid}}$ and
$\mathbb{C}_{T,q}^{\NonSinglet,{\rm id}}$
(see appendix in A in \cite{Rij96b}) as follows
\begin{eqnarray}
  \label{eq:16}
  \lefteqn{
    \mathbb{C}_{A,q}^{\NonSinglet,{\rm nid},(2)}
    - \mathbb{C}_{T,q}^{\NonSinglet,{\rm nid},(2)} =
    \left(\frac{\alpha_s}{4\pi}\right)^2\BigLeftHook
    C_F^2\BigLeftBrace\BigLeftHook 2(1-z)(2\ln z - 4\ln(1-z) + 1)\BigRightHook\cdot}
  \nonumber\\[2ex]
  && \cdot\ln\frac{Q^2}{\mu^2}+ 4(1-z)(4S_{1,2}(1-z) - 8\Li_3(-z)-4\zeta(2)\ln(1-z)
  +4\ln z\Li_2(-z) \nonumber\\[2ex]
  && {}+3\Li_2(1-z)-\ln^2(1-z)-\ln z\ln(1-z)+\frac{25}{2}\ln(1-z))
  + (\frac{24}{5z^2}+\frac{8}{z} \nonumber\\[2ex]
  && {}-16-16z+8z^2+\frac{24}{5}z^3)(\Li_2(-z)
  +\ln z\ln(1+z)) + (-24+8z+8z^2 \nonumber\\[2ex]
  && {}+\frac{24}{5}z^3)\zeta(2) + (10-2z-4z^2
  -\frac{12}{5}z^3)\ln^2 z + (-\frac{24}{5z^2}+\frac{2}{5}+\frac{202}{5}z
  -\frac{24}{5}z^2)\cdot \nonumber\\[2ex]
  && \cdot\ln z + \frac{24}{5z}+\frac{9}{5}-\frac{9}{5}z-\frac{24}{5}z^2
  \BigRightBrace\nonumber\\[2ex]
  && {}+C_AC_F\BigLeftBrace \frac{22}{3}(1-z)\ln\frac{Q^2}{\mu^2} + 4(1-z)(4\Li_3(-z)
  -2S_{1,2}(1-z) \nonumber\\[2ex]
  && {}+2\zeta(2)\ln(1-z)-2\ln z\Li_2(-z)-\frac{25}{6}\ln(1-z))
  +(-\frac{12}{5z^2}-\frac{4}{z}+8+8z \nonumber\\[2ex]
  && {}-4z^2-\frac{12}{5}z^3)(\Li_2(-z)+\ln z\ln(1+z))
  +(4+4z-4z^2-\frac{12}{5}z^3)\zeta(2) \nonumber\\[2ex]
  && {}+ (-2-2z+2z^2+\frac{6}{5}z^3)\ln^2 z
  + (\frac{12}{5z}+\frac{182}{15}-\frac{398}{15}z+\frac{12}{5}z^2)\ln z-\frac{12}{5z}
  \nonumber\\[2ex]
  && {}-\frac{823}{45}+\frac{823}{45}z+\frac{12}{5}z^2\BigRightBrace\nonumber\\[2ex]
  && {}+n_fC_FT_f\BigLeftBrace -\frac{8}{3}(1-z)\ln\frac{Q^2}{\mu^2} -\frac{8}{3}(1-z)
  (\ln(1-z)+\ln z-\frac{19}{6})\BigRightBrace\BigRightHook,\\[2ex]
  \label{eq:17}
  \lefteqn{
    \mathbb{C}_{A,q}^{\NonSinglet,{\rm id},(2)} -
    \mathbb{C}_{T,q}^{\NonSinglet,{\rm id},(2)} =
    \left(C_F^2-\frac{1}{2}C_AC_F\right)
    \left(\frac{\alpha_s}{4\pi}\right)^2\BigLeftHook 8(1+z)(4S_{1,2}(-z)
    -2\Li_3(-z)} \nonumber\\[2ex]
  && {}+ 4\ln(1+z)\Li_2(-z)+2\zeta(2)\ln(1+z)+2\ln z\ln^2(1+z)
  -\ln^2z\ln(1+z) \nonumber\\[2ex]
  && {}-2\zeta(3)) + (\frac{24}{5z^2}-\frac{8}{z}-8z^2+\frac{24}{5}z^3)
  (\Li_2(-z)+\ln z\ln(1+z)) + (8-8z \nonumber\\[2ex]
  && {}-8z^2+\frac{24}{5}z^3)\zeta(2)
  + (-4+4z+4z^2-\frac{12}{5}z^3)\ln^2z+ (-\frac{24}{5z}+\frac{72}{5} \nonumber\\[2ex]
  && {}+\frac{72}{5}z
  -\frac{24}{5}z^2)\ln z+\frac{24}{5z}+\frac{104}{5}-\frac{104}{5}z-\frac{24}{5}z^2
  \BigRightHook,
\end{eqnarray}
where the Riemann zeta-function $\zeta(n)$ and the polylogarithms $\Li_n(z)$, $S_{n,p}(z)$
can be found in \cite{Lew83,*Bar72,*Dev84}.\\
Notice that in $\mathbb{C}_{k,q}^{\NonSinglet,{\rm id}}$ ($k=T,~A$)
we have omitted contributions as
represented by the cut graphs in fig.~\ref{fig:1}. The photon cannot couple to the
cut fermion triangle because of charge conjugation invariance. However the $Z$-boson
decouples too if one sums over all flavours in one family. This is because the $Z$
is connected to the quarks via the axial-vector coupling constant representing
the weak isospin component $I_z^{(f)}$ of a specific flavour $f$ with the property
$\sum_{f=u,d} I_z^{(f)}=0$. From now on we will assume that in the inclusive
state one sums over all members in one family so that the contribution due to
fig.~\ref{fig:1} can be dropped.\\
The first quantity we would like to study is the flavour asymmetry sum rule which is
defined in eq. (2.23) of \cite{Nas94}. It is given by
\begin{eqnarray}
  \label{eq:18}
  \lefteqn{
    \Sigma_A^Q = \sum_{H,f}\,A_f(Q^2)\,\int_0^1dz_1\,\,Q_H^{(f)}\,\BigLeftParen D_f^H(z_1,\mu^2)
    - D_{\bar{f}}^H(z_1,\mu^2) \BigRightParen\cdot}\nonumber\\[2ex]
  && \cdot\int_0^1dz_2\,\,
  \mathbb{C}_{A,q}^\NonSinglet(z_2,Q^2/\mu^2),
\end{eqnarray}
where $Q_H^{(f)}$ is a conserved additive quantity. The first moment of the non-singlet
coefficient function calculated up to order \alphastwo~is equal to
\begin{eqnarray}
  \label{eq:19}
  \lefteqn{
    \int_0^1dz_2\,\mathbb{C}_{A,q}^\NonSinglet(z_2,Q^2/\mu^2) = 1 -
    \left(\frac{\alpha_s(Q^2)}{4\pi}
    \right)^2\,\BigLeftHook 12\beta_0 C_F \zeta(3)\BigRightHook
    + \left(\frac{\alpha_s(Q^2)}{4\pi}\right)^3\BigLeftHook c_{A,q}^{(3)}\BigRightHook,}
  \nonumber\\[2ex]
  &&
\end{eqnarray}
where $\beta_0$ is the lowest order coefficient of the beta-function given by
\begin{equation}
  \label{eq:20}
  \beta_0 = \frac{11}{3}C_A - \frac{4}{3}T_f n_f.
\end{equation}
Notice that the first moments of $\mathbb{C}_{A,q}^\NonSinglet$ and $D_f^H-D_{\bar{f}}^H$ are
separately scheme independent. Further there are no order \alphas~corrections to the
first moment of $\mathbb{C}_{A,q}^\NonSinglet$ \cite{Nas94} and the order \alphastwo~correction
is proportional to $\beta_0$. A comparison between \eref{eq:19} and the order
\alphastwo~corrected $R_{ee} = \sigma_{\rm tot}(e^+e^-\rightarrow X)/\sigma^{(0)}$
\cite{Gor91,*Sur91} reveals that the coefficients of the Riemann zeta-functions
(here $\zeta(3)$ only) are exactly the same. Furthermore if one drops all rational
numbers in $R_{ee}$ one obtains exactly \eref{eq:19}. Following the arguments in
\cite{Bro93,*Gab95} one can make an interesting conjecture  about the third order
term $c_{A,q}^{(3)}$ which has not been calculated yet. Suppose that all rational numbers
in $c_{A,q}^{(3)}$ are zero and that the coefficients of the Riemann zeta-functions
$\zeta(n)$ (here $\zeta(3)$ and $\zeta(5)$) are the same as in $R_{ee}$
then we can make the following conjecture
\begin{eqnarray}
  \label{eq:21}
  \lefteqn{
    c_{A,q}^{(3)} = C_AC_F^2\BigLeftHook -572\zeta(3)+880\zeta(5)\BigRightHook
    + C_FC_A^2\BigLeftHook -\frac{10948}{9}\zeta(3) - \frac{440}{3}\zeta(5)\BigRightHook
    }\nonumber\\[2ex]
  && {}+ C_F^2T_fn_f\BigLeftHook 304\zeta(3)-320\zeta(5)\BigRightHook + C_AC_FT_fn_f
  \BigLeftHook\frac{7168}{9}\zeta(3)+\frac{160}{3}\zeta(5)\BigRightHook
  \nonumber\\[2ex]
  && {}+ C_FT_f^2n_f^2\BigLeftHook-\frac{1216}{9}\zeta(3)\BigRightHook
  + \frac{n_f}{N}d_{abc}d_{abc}\BigLeftHook -8\zeta(3)\BigRightHook,
\end{eqnarray}
where $d_{abc}$ denote the structure constants which emerge from the anti commutation
relations of the generators of the group $SU(N)$.
We now want to study the effect of the order \alphastwo~correction on the asymmetric
fragmentation function and compare the result with the OPAL data \cite{Ake95}. The
fragmentation functions $F_k^H$ will be defined by (see \cite{Ake95})
\begin{equation}
  \label{eq:22}
  F_k^H(x,Q^2) = \frac{1}{\sigma_{\rm tot}}\frac{d\sigma_k^H(x,Q^2)}{dx},\hspace*{8mm}
  (k=T,L,A).
\end{equation}
If we sum over all hadrons of species $H$ we obtain the quantities
\begin{eqnarray}
  \label{eq:23}
  \lefteqn{
    F_k(x,Q^2) = \sum_H\,F_k^H(x,Q^2),\hspace*{8mm}(k=L,T),}\\[2ex]
  \label{eq:24}
  \lefteqn{
    F_A(x,Q^2) = \sum_H\,Q_H\,F_A^H(x,Q^2),}
\end{eqnarray}
where the sum in \eref{eq:24} is taken over all charged hadrons. From \eref{eq:6} and
\eref{eq:23} we infer that $F_A$ gets only contributions from the valence fragmentation
densities $D_{V,f}^H=D_f^H-D_{\bar{f}}^H$. Hence the measurement of $F_A$ provides us with
information about the $x$-behaviour of the valence fragmentation functions. The latter are
available for $H=\pi^\pm, K^\pm, p, \bar{p}$ in \cite{Bin95} where they are parametrized in
leading log (LL) and in next-to leading log (NLL, \MS-scheme). For $H=\pi^+,~K^+,~p$
we obtain from \eref{eq:6} and \eref{eq:22}
\begin{eqnarray}
  \label{eq:24_2}
  \lefteqn{
    F_A^H(x,Q^2) = \frac{1}{\sigma_{\rm tot}}\,\int_x^1\frac{dz}{z}\,\BigLeftHook
    A_U(Q^2)\,D_{V,U}^H\left(\frac{x}{z},\mu^2\right) - A_D(Q^2)\,D_{V,D}^H\left(
    \frac{x}{z},\mu^2\right)\BigRightHook\cdot} \nonumber\\[2ex]
  && \cdot\mathbb{C}_{A,q}^\NonSinglet(z,Q^2/\mu^2),
\end{eqnarray}
with $U=u$ ($\pi^+,K^+$) and $D=d$ ($\pi^+$) or $D=s$ ($K^+$). The proton contribution
is given by $F_A^P = 0.16~F_A^{\pi^+}$ where the factor 0.16 originates from
\cite{Bin95} where one has estimated $F^{p+\bar{p}} = F_L^{p+\bar{p}} + F_T^{p+\bar{p}}$
by putting $F^{p+\bar{p}}=0.16~F^{\pi^++\pi^-}$. Since in \cite{Bin95} one has taken
$D_{V,U}^H = D_{V,D}^H$ we observe that $F_A^H(x,M_Z^2)$ \eref{eq:24_2} is negative over the
whole $x$-region. This property can be traced back to the value of the electroweak angle
leading to $A_D(M_Z^2)/A_U(M_Z^2)\sim 2$. Therefore all hadrons with positive charge
($Q_H > 0$) give a negative contribution to $F_A(x,M_Z^2)$ \eref{eq:24}. If $\bar{H}$ is the
anti-particle of $H$ we have the relation $F_A^{\bar{H}}=-F_A^H$. Because of $Q_{\bar{H}} = -Q_H$
in \eref{eq:24} the anti-particles ($\pi^-, K^-, \bar{p}$) also give a negative contribution to
$F_A(x,Q^2)$. Therefore the parametrization in \cite{Bin95} predicts a negative $F_A(x,Q^2)$
\eref{eq:24} over the whole $x$-region at $Q^2=M_Z^2$.\\
In our plots discussed below a comparison will be made with the OPAL data \cite{Ake95} so that
we have to choose $Q^2=M_Z^2$. Further we take $\mu^2=Q^2$ in \eref{eq:6} and
$n_f=5$. The running coupling constant is chosen to be $\alpha_s(M_Z^2) = 0.126$.
Finally we want to emphasize that a full next-to-next-to-leading (NNLO) analysis
of $F_T$ and $F_A$ is not possible yet because of the missing three-loop contributions
to the DGLAP splitting functions. Therefore the order \alphastwo~correction, which can
be only attributed to the coefficient functions in \eref{eq:16}, \eref{eq:17} and
\cite{Rij96b}, have to be considered as an estimate. The order \alphastwo~corrected $F_L$
is complete because here the NLL fragmentation densities and the order \alphastwo~
corrected coefficient functions are available (see \cite{Rij96}).\\
In fig.~\ref{fig:2} we have plotted $F_A^{LO}$, $F_A^{NLO}$ and $F_A^{NNLO}$ together
with the OPAL data (see also fig. 4 in \cite{Ake95}). There is a difference between
$F_A^{LO}$ and $F_A^{NLO}$ but the order \alphastwo~corrections shown by $F_A^{NNLO}$ are
unobservable. Furthermore the theoretical curves are above the data. In fig. 8
of \cite{Ake95} the OPAL-collaboration also presented the data for the ratio
\begin{equation}
  \label{eq:25}
  R_A(x,Q^2) = \frac{F_A(x,Q^2)}{F(x,Q^2)}, \hspace{8mm}F(x,Q^2) = F_T(x,Q^2) = F_L(x,Q^2).
\end{equation}
In fig.~\ref{fig:3} these data are compared with the theoretical predictions
$R_A^{LO}$, $R_A^{NLO}$, and $R_A^{NNLO}$. Here we see the same features as has been
observed for $F_A$ in \fref{fig:2}. There is
no difference between $R_A^{NLO}$ and $R_A^{NNLO}$ and only the order \alphas~corrections,
represented by $R_A^{NLO}$, are visible. Also in this case the data are below the theoretical
predictions.\\
From the data one can infer the integrated fragmentation function for which the
theoretical predictions corrected up to order \alphastwo~are given below
\begin{eqnarray}
  \label{eq:26}
  \lefteqn{
    \int_{0.1}^1dx\,F_A^{NNLO}(x,M_Z^2) = -0.016 \hspace*{3mm}(-0.023),}\\[2ex]
  \label{eq:27}
  \lefteqn{
    \int_{0.1}^1dx\,\frac{1}{2}x\,F_A^{NLO}(x,M_Z^2) = -0.0020 \hspace*{3mm}(-0.0027).}
\end{eqnarray}
The experimental values for \eref{eq:26} and \eref{eq:27} are -0.0229$\pm$ 0.0044
and -0.00369$\pm$ 0.00046
respectively. Since the fragmentation densities in \cite{Bin95} have a limited range
of validity we have imposed a lower bound on the integration which is given by $x=0.1$.
Between the brackets in \eref{eq:26}, \eref{eq:27} we have quoted the LO results. It
turns out that the latter are in better agreement with experiment than the NLO
and NNLO numbers. Further the values of the integrals also hold in NLO since the order
\alphastwo~corrections are extremely small.\\
The OPAL-data indicate that at low $x$ $F_A(x,M_Z^2)$ might become positive. If this
is the case one has to assume that in this region $D_{V,U}^H(x,\mu^2) > D_{V,D}^H(x,\mu^2)$
provided the zeroth order contribution to $\mathbb{C}_{A,q}^\NonSinglet$ which is given
by $\delta(1-z)$ dominates the integral.\\
Summarizing the above we conclude that the order \alphastwo~corrections to $F_A$ are
negligible and we do not expect that this will change when the effect of the
three-loop DGLAP splitting functions are taken into account. Furthermore the above
results reveal that the measurement of $F_A$ puts some constraints on the valence
fragmentation densities. In particular it means that the NLL parametrizations in
\cite{Bin95} have to be modified in order to get agreement with the OPAL data.

%
\bibliographystyle{/home/rulil0/pieter/tex/style/mybib}
\bibliography{/home/rulil0/pieter/tex/physics}

\newpage
\begin{mcbibliography}{10}
\bibitem{Bra90}{W. Braunschweig \etal~(TASSO),}{Z. Phys. {\bf C47} (1990)
  187}\bibitem{Pet88}{A. Peterson \etal~(Mark II),}{Phys. Rev. {\bf D37} (1988)
  1}\bibitem{Aih88}{H. Aihara \etal~(TPC/2$\gamma$),}{Phys. Rev. Lett. {\bf 61}
  (1988) 1263}\bibitem{Pod}{O. Podobrin (CELLO),}{Ph. D. thesis, Universit\"at
  Hamburg}\bibitem{Li90}{Y.K. Li \etal~(AMY),}{Phys. Rev. {\bf D41} (1990)
  2675}\bibitem{Abr93}{P. Abreu \etal~(DELPHI),}{Phys. Lett. {\bf B311} (1993)
  408}\bibitem{Bus95}{D. Buskulic \etal~(ALEPH),}{Phys. Lett. {\bf B357} (1995)
  487}\bibitem{Ake95}{R. Akers \etal~(OPAL),}{Z. Phys. {\bf C68} (1995)
  203}\bibitem{Nas94}{P. Nason, and B.R. Webber,}{Nucl. Phys. {\bf B421} (1994)
  473}\bibitem{Bai79}{R. Baier, and K. Fey,}{Z. Phys. {\bf C2} (1979)
  339}\bibitem{Alt79b}{G. Altarelli, R.K. Ellis, G. Martinelli, and S.-Y.
  Pi,}{Nucl. Phys. {\bf B160} (1979) 301}\bibitem{Cur80}{G. Curci, W.
  Furmanski, and R. Petronzio,}{Nucl. Phys. {\bf B175} (1980)
  27}\bibitem{Fur80}{W. Furmanski, and R. Petronzio,}{Phys. Lett. {\bf 97B}
  (1980) 437}\bibitem{Flo81}{E.G. Floratos, C. Kounnas, and R. Lacaze,}{Nucl.
  Phys. {\bf B192} (1981) 417}\bibitem{Rij96}{P.J. Rijken and W.L. van
  Neerven,}{INLO-PUB-4/96, hep-ph/9604436, to be published in Phys. Lett.
  B}\bibitem{Rij96b}{P.J. Rijken and W.L. van Neerven,}{INLO-PUB-09/96,
  hep-ph/9609377}\bibitem{Gor91}{S.G. Gorishni, A.L. Kataev, and S.A.
  Larin,}{Phys. Lett. {\bf 159B} (1991) 144}\bibitem{Sur91}{L.R. Surguladze,
  and M.A. Samuel,}{Phys. Rev. Lett. {\bf 66} (1991) 560, Erratum Phys. Rev.
  Lett. {\bf 66} (1991) 2416}\bibitem{Hoo72}{G. 't Hooft and M. Veltman,}{Nucl.
  Phys. {\bf B44} (1972) 189}\bibitem{Bre77}{P. Breitenlohner and B.
  Maison,}{Common. Math. Phys. {\bf 53} (1977) 11, 39, 55}\bibitem{Aky73}{D.
  Akyeampong and F. Delbourgo,}{Nuovo Cim. {\bf 17A} (1973)
  578}\bibitem{Aky73b}{D. Akyeampong and F. Delbourgo,}{Nuovo Cim. {\bf 18A}
  (1973) 94}\bibitem{Aky74}{D. Akyeampong and F. Delbourgo,}{Nuovo Cim. {\bf
  19A} (1974) 219}\bibitem{Lar93}{S.A. Larin,}{Phys. Lett. {\bf 303B} (1993)
  113}\bibitem{Lar94c}{S.A. Larin,}{Phys. Lett. {\bf 334B} (1994)
  192}\bibitem{Zij92}{E.B. Zijlstra, and W.L. van Neerven,}{Nucl. Phys. {\bf
  B383} (1992) 525}\bibitem{Lew83}{L. Lewin,,}{``Polylogarithms and Associated
  Functions'' (North Holland, Amsterdam 1983)}\bibitem{Bar72}{R. Barbieri, J.A.
  Mignaco, and E. Remiddi,}{Nuovo Cim. {\bf 11A} (1972) 824}\bibitem{Dev84}{A.
  Devoto, and D.W. Duke,}{Riv. Nuovo Cim. {\bf 7-6} (1984)
  1}\bibitem{Bro93}{D.J. Broadhurst and A.L. Kataev,}{Phys. Lett. {\bf 315B}
  (1993) 179}\bibitem{Gab95}{G.T. Gabadadze and A.L. Kataev,}{JETP Lett. {\bf
  61} (1995) 448}\bibitem{Bin95}{J. Binnewies, B.A. Kniehl, and G. Kramer,}{Z.
  Phys. {\bf C65} (1995) 471}\end{mcbibliography}
\newpage
\section{Figure captions}
{\bf Fig. 1} Diagrams with four quarks in the final state containing a cut triangular
             quark loop.\\[2ex]
{\bf Fig. 2} Contributions to the asymmetry fragmentation function $F_A(x,Q^2)$ \eref{eq:24}
             at $Q=M_Z$ using the fragmentation density set of \cite{Bin95}. Solid line: LO.
             Dashed line: NLO. Dotted line: NNLO. The exprimental data are taken from
             OPAL \cite{Ake95}.\\[2ex]
{\bf Fig. 3} The ratio $R_A(x,Q^2)$ \eref{eq:25} at $Q=M_Z$ using
             the fragmentation density set of \cite{Bin95}. Short dashed line: LO.
             Solid line: NLO. Dotted line: NNLO. The exprimental data are taken from
             OPAL \cite{Ake95}.
%

\newpage
%
%
%
%
%
%
\clearpage
%

%
\end{document}